# Angle-resolved Photoemission Spectroscopy

Hongyun Zhang[1], Tommaso Pincelli[2], Chris Jozwiak[3], Takeshi Kondo[4,5], Ralph Ernstorfer[2,6], Takafumi Sato[7,8], Shuyun Zhou[1,9,*]

[1] State Key Laboratory of Low-Dimensional Quantum Physics and Department of Physics, Tsinghua University, Beijing 100084, People's Republic of China

[2] Fritz Haber Institute of the Max Planck Society, Faradayweg 4-6, 14195 Berlin, Germany

[3] Advanced Light Source, Lawrence Berkeley National Laboratory, Berkeley, CA 94720, USA

[4] Institute for Solid State Physics, The University of Tokyo, Kashiwa, Chiba 277-8581, Japan

[5] Trans-scale Quantum Science Institute, The University of Tokyo, Bunkyo-ku, Tokyo 113-0033, Japan

[6] Institut für Optik und Atomare Physik, Technische Universität Berlin, Straße des 17. Juni 135, 10623 Berlin, Germany

[7] Advanced Institute for Materials Research (WPI-AIMR), Tohoku University, Sendai 980-8577, Japan

[8] Department of Physics, Graduate School of Science, Tohoku University, Sendai 980-8578, Japan

[9] Frontier Science Center for Quantum Information, Beijing 100084, People's Republic of China

[*] Email: syzhou@mail.tsinghua.edu.cn

**For solid-state materials, the electronic structure, *E(k)*, is critical in determining a crystal's physical properties. By experimentally detecting the electronic structure, the fundamental physics can be revealed. Angle-resolved photoemission spectroscopy (ARPES) is a powerful technique for directly observing the electronic structure with energy- and momentum-resolved information. Over the past decades, major improvements in the energy and momentum resolution, alongside the extension of ARPES observables to spin (SpinARPES), micrometer or nanometer lateral dimensions (MicroARPES/NanoARPES), and femtosecond timescales (TrARPES), have led to major scientific advances. These advantages have been achieved across a wide range of quantum materials, such as high-temperature superconductors, topological materials, two-dimensional materials and heterostructures. This primer introduces key aspects of ARPES principles, instrumentation, data analysis, and representative scientific cases to demonstrate the power of the method. Perspectives and challenges on future developments are also discussed.**

**Introduction**

Angle-resolved photoemission spectroscopy (ARPES) is a powerful technique for directly probing the energy- and momentum-resolved electronic structure, which is fundamental in condensed matter physics[1,2]. The basic principle of ARPES is the photoelectric effect, where photoelectrons are emitted following irradiation with monochromatic light, which was first observed by H. Hertz[3] and explained by A. Einstein[4]. By resolving the energy and emission angle of photoelectrons, ARPES can provide critical information about the electronic structure, and has played a leading role in revealing the fundamental physics of solid-state materials[1,2].

ARPES is often succinctly described as a photon-in, electron-out experiment. Monochromatic light with photon energy ($hv$) larger than the material's work function ($\Phi$) is focused onto the sample surface, and photoelectrons are emitted[3,4]. A spectrometer records the intensity distribution of the photoelectrons as a function of kinetic energy ($E_k$) and the emission angle α+β, Fig. 1a top panel. The resulting photoelectron spectrum collected by an electron analyzer (Fig. 1a), $I(E_k, α+β)$, directly reflects the electronic structure $E(k)$, namely, the energy ($E$) *vs.* momentum ($k$) relation of electrons inside the solid-state material as illustrated in Fig. 1a top panel. Such intensity map can be used to map out the three-dimensional (3D) electronic structure (Fig. 1b). This type of *k*-space electronic structure image is typically observable in real-time, with images acquired over seconds to hours, depending on experimental specifics and statistics desired.

The strong link between the measured photoelectron spectrum and the material's electronic structure is due to the conservation of energy and in-plane momentum — parallel to

the sample surface — in the photoemission process[1]. The conservation of energy is expressed as:

$$E_k = h\nu - \Phi - E_B \text{ (eq. 1)}$$

where $E_k$ is the photoelectron kinetic energy and $E_B$ is the corresponding electronic state binding energy with respect to the Fermi level, $E_F$, as is schematically illustrated in Fig. 1c. The photoelectron's momentum value $|\mathbf{k}|$ is given by:

$$|\mathbf{k}| = 1/\hbar \sqrt{2m_e E_k} \text{ (eq. 2)}$$

Where $\hbar$ is the Planck's reduced constant, and $m_e$ is the electron mass. The in-plane momentum components, $k_x$ and $k_y$, are conserved in the photoemission process due to the translational symmetry of the crystal along these directions. This provides a direct mapping of emission angles to corresponding electronic states' momenta. For the geometry shown in Fig. 1a, for example, $k_x$ can be approximated by

$$k_x = \frac{\sqrt{2m_e}}{\hbar} \cdot \sqrt{E_k} \sin(\alpha + \beta) \text{ (eq. 3)}$$

By collecting the data at different sample angles, for example $\theta$ in Fig 1a, a 3D dataset of the electronic structure as a function of $E_B$, $k_x$, and $k_y$ can be acquired (Fig. 1b). Constant energy maps, or ARPES images as a function of $k_x$ and $k_y$ at specific $E_B$ can straightforwardly be extracted from such a dataset and analyzed. The constant energy map at the Fermi level, where $E_B = 0$, represents the Fermi surface (FS) map, Fig. 1b bottom panel.

Over the past decades, ARPES has witnessed major advances in both instrumentation and scientific discoveries. The development of light sources such as 3<sup>rd</sup> generation synchrotrons[5-10], deep ultraviolet (UV) lasers[11-15], high-harmonic generation (HHG) sources[16-21], and the extension to the soft/hard X-ray range[5,9,22,23] have broadened the scope of ARPES. Together with instrumentation improvements, such as reduced beam spot, higher spin detection efficiency, and integration with pump-probe technique, these developments have increased the impact of ARPES. This has enabled analysis of samples with different dimensionalities, physical properties, buried and interfacial electronic states, and multiple electronic degrees of freedom including spin, pseudospin, and valley. For example, high-resolution ARPES (HR-ARPES)[10,12,13,19,24] has played a critical role in revealing the low-energy spectrum of electrons in superconductors and topological materials. The greatly reduced spot size of nano- or micro-spot ARPES (NanoARPES/MicroARPES)[25-31] makes it possible to reveal the electronic structure of small samples with a micrometer (μm) or submicrometer size, as well as phase-separated materials. Moreover, the development of pulsed light sources with femtosecond (fs) or picosecond (ps) pulse duration enables the detection of electronic dynamics in the ultrafast time scale by time-resolved ARPES (TrARPES)[16-18,32-34]. The two-order-of-magnitude improvement in the spin detection efficiency has also extended spin-resolved ARPES (SpinARPES)[35-39] measurements to a much wider range of materials with rich spin-related physics. The application of ARPES has led to major scientific progress in quantum materials[2,40], such as high-temperature superconductors[1,2,41], topological insulators[42,43] and semimetals[44-46], and two-dimensional (2D) materials[47-50] and heterostructures[30,51,52]. For in-depth ARPES reviews, see the book on photoemission and these review articles[1,2,42-44,47,48,50,52].

ARPES measurements have become more accessible to researchers across physics, material science and beyond. This Primer gives a brief introduction of the basic principles of ARPES and its variations including HR-ARPES, SpinARPES, NanoARPES/MicroARPES, TrARPES, SX-ARPES (soft X-ray ARPES), and key aspects of experimentation and data analysis. Representative ARPES scientific cases are presented to illustrate the impact of the technique, revealing the physics of superconductors, topological materials, 2D materials and heterostructures. Possible improvements and future perspectives are also discussed.

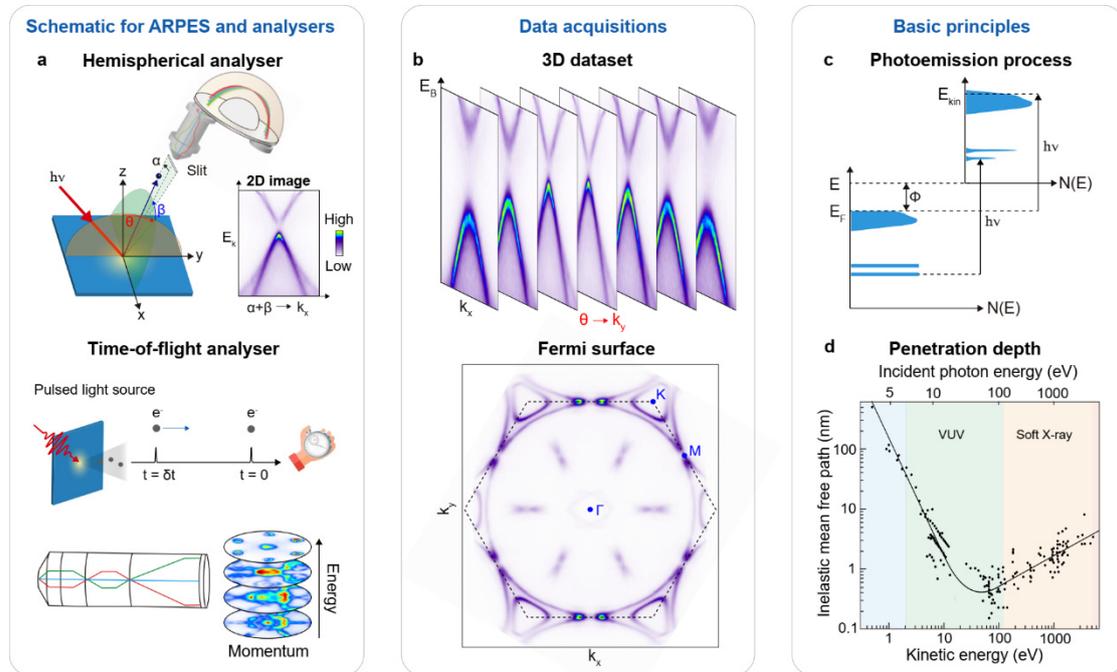

**Fig. 1. ARPES basic principles and experimentation**. (a) Top panel: schematic of the basic ARPES setup with the hemispherical analyzer and an example 2D $E$-$k_x$ data. Bottom panel: schematic illustration of the TOF detector and an example 3D $E$-$k_x$-$k_y$ data. (b) Parallel cuts measured at different sample $\theta$ angles, from which the FS map can be obtained (bottom panels). (c) Schematic illustration of the photoemission process. (d) The universal curve of photoelectrons' inelastic mean free path as a function of $E_k$ (bottom axis) and the incident photon energy (top axis) reproduced from[53].

**Experimentation**

In this section, the fundamental experimental principles of ARPES are introduced, alongside the key components of ARPES systems. This technique is wonderfully rich in details and subtle complexities, yet standard modern setups can readily provide fascinating and impactful results for beginners and experts, alike. This primer provides the basics for an experimentalist. Comprehensive ARPES treatments can be found in ARPES books and review articles[1,2,54].

**Key ARPES components**

The key aspects of any ARPES measurements can be divided into three categories: the light sources, the photoelectron spectrometer, and the sample.

**Light sources**

The light source and its parameters have an enormous impact on ARPES measurements. First, the ARPES spectrum is strongly dependent on the photon energy, or wavelength. The photon energy of the light source determines the accessible range of $E_B$ and $\boldsymbol{k}$ (through equations 1 and 2) and plays a role in determining the out-of-plane momentum, $k_z$, sampled in 3D materials[1]. The photon energy also affects the probing depth of ARPES measurements[53]. Photoelectrons must travel through the material and escape the surface without inelastic collisions to contribute to the detectable ARPES signal. Therefore, the probing depth of ARPES is determined by the inelastic mean free path of electrons in solids, which is typically far less than the penetration length of the light, and follows a general $E_k$ (or effectively $h\nu$) dependence known as the Universal Curve[53], Fig. 1d. The probing depth can vary from a few angstroms (Å) for $h\nu \sim 100$ eV, to several nanometers (nm) for $h\nu > 1$ keV, to near 100 nm for $h\nu < 7$, allowing ARPES measurements to be more surface or bulk sensitive.

ARPES intensity is strongly dependent on both the photon energy and photon polarization due to the dipole matrix elements that modulate the photoemission process[55]. Controlling these parameters can provide useful sensitivity to important characteristics of a given electronic structure. The light source bandwidth — monochromaticity or energy resolution — and brightness are also critical, as they impact the energy resolution of the collected ARPES data and the data acquisition efficiency at which ARPES can achieve sufficiently good statistics, respectively.

There are three major types of light sources for ARPES measurements: synchrotron light sources[5-10], gas (helium, neon, argon, krypton or xenon) discharge lamps[56-58], and continuous-wave (CW) or pulsed lasers[11-21]. These light sources have different characteristics in terms of accessible photon energy, linewidth and timing structures. The linewidth affects the resolution, while timing structures relate to the detection scheme and time resolution. A comparison of those light sources is listed in Table 1.

Among these light sources, third generation synchrotron light sources provide nearly ideal performance, with high brightness, wide and continuously tunable energy ranges, tunable resolution-versus-intensity trade-offs, and selectable polarization[5-9]. Beamlines commonly cover various useful photon energy ranges within 5 eV to several keV, and can now provide resolving powers, where resolving power = photon energy/energy resolution, of 30,000 or more[59]. In addition, they provide energy resolutions of nearly 1 meV at the lower photon energy ranges, such as 30 eV.

While synchrotrons are ideal light sources in many ways, other light sources are useful for ARPES measurements and can excel in specific ways. Small noble gas discharge lamps provide bright, high-resolution sources of fixed, specific energy lines that enable high-quality, laboratory-based ARPES measurements.

| Light source | Parameters | Advantages | Disadvantages |
| --- | --- | --- | --- |
| 3rd generation synchrotron light[5-9] | Widely tunable $h\nu$ from < 5 eV to several keV; Photon flux of ~ $10^{12} - 10^{13}$ photons/s; Typical spot size ~ 10 - 100 μm (50~150 nm for NanoARPES). | High brightness; Selectable polarization control; Tunable surface (VUV) and bulk sensitivity (soft-X ray). | User endstations at large scale facilities; Mostly without fs time resolution. |
| Gas discharge lamps[56-58] | Discrete $h\nu$ determined by the energy lines of discharge gases (helium, neon, argon, krypton, xenon); Photon flux of ~ $10^{12}$ photons/s; Typical spot size ~ 0.3 - 1 mm. | Capability of compact, stable, and economical Lab-based ARPES setups; High energy resolution. | Relatively low photon flux at fixed $h\nu$, and large beam spot; Mostly lack of polarization and time resolution. |
| Table-top lasers[11-21,60] | Discrete or narrowly tunable $h\nu$ within 5.3 – 7 eV[14,28,60], 11 eV or higher harmonic generations within 8 – 40 eV [16-21]; Photon flux of ~ $10^9 – 10^{15}$ photons/s; Typical spot size < 100 μm (several μm for MicroARPES). | High brightness and high energy resolution for low $h\nu$ (< 7 eV)[12,13,19]; Capability of pump-probe experiments with fs or ps time resolution[16-18,32-34], selectable or resonant excitations by tuning the pump light. | Low $h\nu$ range (< 7 eV) limits the detectable energy and $k$-space range, while HHG source has limited flux or resolution. |

Table 1. Comparison of different light sources.

Over the past decade, table-top laser sources have become successful and popular through the harmonic generation in nonlinear crystals[33] or HHG systems using gas as a nonlinear medium[16-21] to obtain photon energies high enough for ARPES measurements. Such systems provide some amount of tunability, polarization control[11-13,60], and extremely high energy resolution[12,13,19]. Many are ultrafast light sources[16-18,32-34], with ultrashort pulses at fs time scales, enabling the new stroboscopic technique of TrARPES measurements. The scientific interests have also resulted in the development of TrARPES with the latest generation of high repetition rate free-electron lasers[61].

**Photoelectron Spectrometer**

Basic ARPES data encompass a 3D phase space: $E_k$ and 2 angular dimensions that constitute the possible $2\pi$ hemisphere of the free space above the sample surface. There are various spectrometer designs for data acquisition, while two are currently favored and commercially available including the hemispherical analyzer[62,63] (Fig. 1a) and the time-of-flight (TOF) analyzer[64-67] as illustrated in Fig. 1a bottom panel.

The hemispherical analyzer is one of the most widespread and offers the highest commercial refinement. It consists of a cylindrical electrostatic lens, a hemispherical capacitor, and an electron detector. Photoelectrons emitted within a certain angular range (typically, $\alpha$ is about ± 15°) along the lens's optical axis enter the lens and are focused onto a linear slit at the entrance of the hemisphere, which selects out photoelectrons along a single angular axis $\beta$. The hemisphere capacitor forms a spherical potential that disperses the photoelectrons according to their $E_k$. Photoelectrons with lower $E_k$ follow tighter orbits around the hemisphere, while photoelectrons with higher $E_k$ follow larger orbits[62,63]. An electron detector, typically formed by a multichannel plate electron multiplier (MCP), a phosphor screen, and a digital camera, records a 2D photo-electron image at the exit of the hemisphere, illustrated by the 2D intensity map in Fig. 1a. This image directly corresponds to a slice through the 3D phase space, with one axis corresponding to $E_k$, and the other axis to the emitting angle $\alpha+\beta$, which can be directly transformed into a map of $E_B$ vs. $k_x$. The full electronic structure in the 3D $E_B$-$k_x$-$k_y$ phase space can be obtained by sequentially scanning the other angular dimension by rotating the relative orientation ($\theta$) between the sample and spectrometer as shown in Fig. 1a, or with electrostatic deflectors in the lens, acting perpendicularly to the analyzer slit.

The TOF analyzer resolves photoelectron kinetic energy by measuring the transit time from the sample to the final electron detector[64-67], Fig. 1a bottom panel. This detection scheme requires pulsed light sources of a particular timing structure and a time-resolved electron detector. Typically, the time resolutions of the pulses are on the order of 100 picoseconds (ps) or better. The electron detector resolves both the 2D position and arrival time of each incident photoelectron, enabling the TOF analyzer to acquire full 3D $E_B$-$k_x$-$k_y$ ARPES datasets in parallel without the need to sequentially scan the second angle, or $k$ axis. While this setup places restrictions on the light sources used, it provides significantly improved data collection efficiency when full $k_x$-$k_y$ information is desired and is particularly useful for applications with a low photon flux, for example, when using HHG light sources.

In addtion to the hemispherical and TOF analyzers, a new concept has been developed recently: the momentum microscope style spectrometer[68,69]. This type of spectrometer has the ability to acquire data through both in-plane $k$ directions simultaneously across a wide range of momentum space. The lens is placed very close to the sample (∼ 1 mm) with the sample oriented perfectly perpendicular to the optical axis. A strong electrical extraction voltage (several to tens of kV) is applied between the sample and lens, pulling the entire 2π angular range of photoelectrons from the upper hemisphere into the lens. Photoelectron energy can be resolved in these spectrometers with the inclusion of hemispheres and apertures to filter out electrons of particular energies[68,69]. A final electron detector then records a 2D $k_x$-$k_y$ photoelectron image at a certain energy, and the full 3D $E_B$-$k_x$-$k_y$ phase space is obtained by sequentially scanning the filtered energy. A useful aspect of this style of instrument is that it can readily tune the lens to switch between a real-space and $k$-space image at the detector, enabling both ARPES measurements and photoemission electron microscopy (PEEM) measurements[68]. TOF-based momentum-microscopes have also been developed, bringing similar enhancements to 3D phase-space collection efficiency as above[61,69-71]. Each of these different types of detectors has its advantages and disadvantages, and a quantitative comparison of TOF momentum microscope and the hemispherical analyzer can be found in this article[72].

**Sample**

The sample is a critical component of any ARPES experiment. It must be conductive enough to drain the total induced photocurrent, without resulting in the sample charging to a disruptive voltage that affects the energy position as well as the peak width. The total photocurrent can vary significantly across the range of possible ARPES experiments, but in practice are typically in the range of nanoamps (nA). From a simple Ohm's law model, if the sample has a total resistance of less than ∼100 kΩ to the experimental ground, the resulting induced voltage can be kept to less than 1 meV. Samples that are strong insulators are very challenging to measure with ARPES, particularly at low temperatures. Experimentally, a simple way to test whether there is sample charging is to vary the photon flux, which controls the total photocurrent, and checks if the ARPES spectrum shifts in energy[73]. Preparation and mounting strategies to ensure good electrical conductivity can be used to mitigate issues of sample charging. For extremely insulating samples, increasing the sample temperature to 100 or 150 K are helpful to reduce the sample charging via thermal population of the conduction band. There are also other strategies to populate the conduction band artificially. Short-wavelength (400 nm), large-focus CW lasers are used to induce sufficient conductibility to drain the photoemission current. Low-energy electron guns, known as flood guns, are also an option to compensate the photoemitted charge by injecting electrons into the solid.

Since ARPES is a highly surface-sensitive technique, the sample must also have a very clean surface, free of contaminants and adsorbents. While sample surface chemistry can differ, reactive samples must be kept in ultra-high vacuum (UHV) in the low $10^{-11}$ torr range to maintain a clean and consistent surface for the 10 to 20 hours of a typical ARPES experiment[74]. UHV conditions are always preferred even for relatively inert surfaces due to physisorption.

The sample preparation is critical for ARPES experiments, including two key aspects: an atomically clean surface and good electrical contact between the sample and the ground. For single crystals with well-defined cleaving planes, the most common way to obtain a clean surface is to cleave[75] the sample inside the vacuum chamber as shown in the left two panels in Fig. 2a. Before the samples are loaded into the ARPES chamber, a cleaving bar — for example, a ceramic post — is glued to the sample surface using conductive silver epoxy glue[76] or Torr seal with a stronger cleaving force[77]. There are different types of glues with different cleaving strengths and curing temperatures[76], the glue used should be chosen accordingly. In the case of insulating glue, conductivity paint or graphite is applied afterward to ensure good electrical conductivity.

For samples that cannot be cleaved, the sample surface can be mechanically polished, then cleaned *in situ* by sputtering and annealing cycles[78]. A clean surface can also be obtained by simple *in situ* degassing and annealing to remove adsorbed molecules, which has successfully been applied to inert thin-film samples and thin flakes exfoliated onto a substrate. The annealing temperature is generally limited by the decomposition temperature of the films. For example, for epitaxial graphene, annealing at 250 to 400 ℃ is sufficient to expose a very clean surface. For some materials, however, the only choice is *in situ* synthesis, where the samples are produced and measured without breaking UHV. ARPES is naturally compatible with *in situ* growth methods such as molecular beam epitaxy (MBE)[79], which enables production of very high-quality sample surfaces, with precise layer and stoichiometry control and may even be used for growing chemically unstable or metastable samples[80,81]. Other films and samples can be successfully measured with ARPES by avoiding atmospheric exposure between the synthesis and measurement chambers through the use of integrated or interconnected synthesis and ARPES systems, the use of removable protective capping layers[82,83], and transporting samples in UHV-quality vacuum suitcases.

Samples can be mounted onto the sample holder by conductive epoxy glue, as shown in the left panels of Fig. 2a. Sample holders are often made of copper for efficient heat transfer, mostly for cryogenic temperatures. When the samples are expected to be annealed, glue is generally avoided and spot welding or screws are commonly used, right two panels in Fig. 2a. Annealable sample holders are usually made of refractory materials such as tantalum or molybdenum that do not decompose or outgas even when heated to high temperatures.

During ARPES measurements, *in situ* sample control and manipulation are important to tune the electronic structure and the material properties. Fig. 2b shows three methods to manipulate the electronic structure in ARPES measurements, *in situ* electron doping[84,85], electrostatic gating[86], and strain engineering[87,88]. The *in situ* alkali deposition is usually used to tune the carrier concentration of the sample[84,85]. This can control the electron doping in a large range and sometimes lead to new quasiparticle formation, such as plasmarons[89,90], or new electronic states in crystalline insulator[91]. A slightly different situation occurs when, in layered materials, the alkali atoms intercalate into the van der Waals gap. For example, in Li-intercalated graphene with Kekulé order, extended flat bands and intriguing physics, such as the chiral symmetry breaking, can be realized [92,93]. This suggests the unique capability to induce intercalation by *in situ* metal deposition. To produce a very small coverage of the surface without segregation into islands, the evaporation is often performed at cryogenic temperatures to limit alkali mobility. Another way to tune the carrier density is by electrostatic gating[86] as shown in the middle panel of Fig. 2b. Electrostatic gating can provide more accurate control and reversible doping, without introducing disorders, compared to surface alkali metal deposition. Strain is another sample control method that has been increasingly applied to ARPES measurements[87,88]. By applying *in situ* strain to the samples, intriguing physics such as phase transitions has been realized[87,88]. These *in situ* sample manipulation methods provide new opportunities to tailor the electronic structure during ARPES measurements.

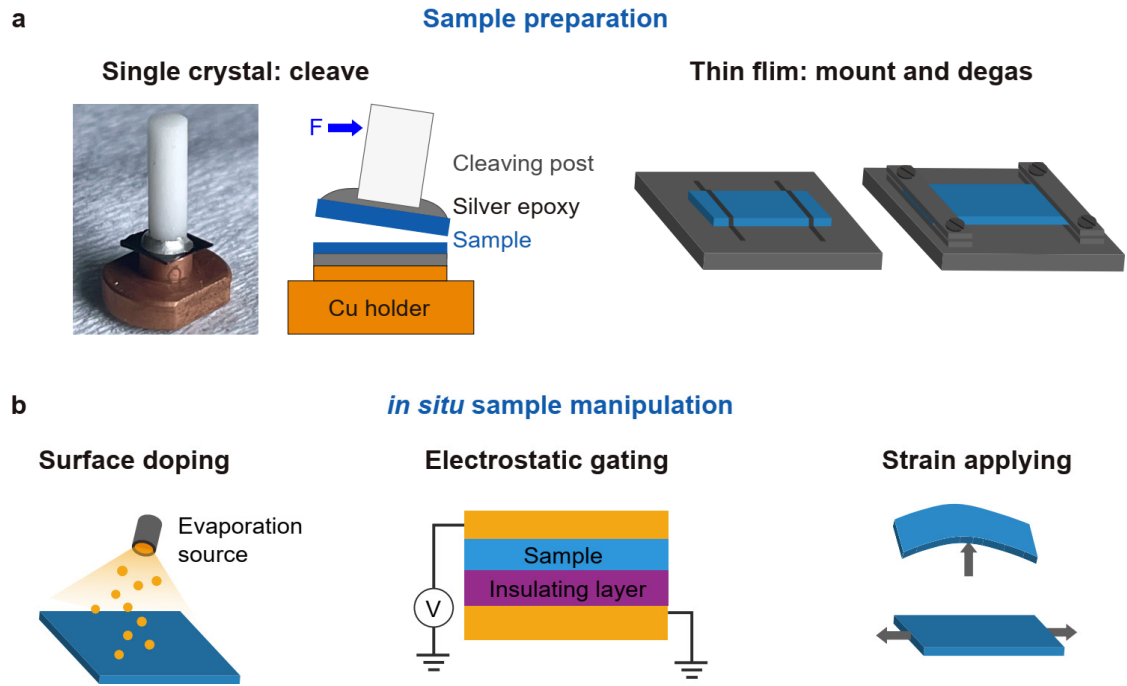

**Fig. 2. Sample preparation and controls in ARPES measurements.** (a) Left two panels: a photo and schematic drawing of the sample preparation using the cleaving method for single crystals; Right two panels: schematic drawings of mounted thin films by spot-welding and screwing for degassing and exposing a clean surface. (b) Schematic drawings of *in situ* surface electron doping, sample manipulation by electrostatic gating, and applying strain onto the sample during ARPES measurement.

**Basic setup and variations**

The instrumentation development of ARPES and its expansion along many directions have led to variations such as HR-ARPES, SX-ARPES, SpinARPES, NanoARPES/MicroARPES, and TrARPES. However, there are some general similarities across the different ARPES variations, and it is instructive to discuss the basic setup first.

**Basic setup**

The basic ARPES setup is illustrated in Fig. 3, consisting of a UHV chamber, an ARPES spectrometer, and a sample stage.

The UHV requirement has a significant impact on the design of ARPES systems because it requires a UHV chamber, vacuum pumps, and the use of bakeable (stable up to 150°C) and UHV compatible materials throughout. Since the momentum of photoelectrons is deduced from the emission angle, the photoelectrons' path to the spectrometer must not be perturbed by uncontrolled fields. This means the area around the sample and spectrometer must be well shielded from magnetic fields, including the Earth's and other sources. Such nearly field-free environment is usually achieved by extensive mu-metal magnetic shielding — yellow and blue layers in the right cutaway view of Fig. 3 — to screen the magnetic fields in the chamber. Additionally, strictly non-magnetic materials should be used inside this shielding.

The ARPES spectrometer shown in Fig. 3 is a standard hemispherical analyzer, which typically represents a significant portion of the setup, in terms of both physical size and cost. As resolution tends to scale better with increasing size, high-performance systems have fairly large analyzers. For example, the energy resolution of the hemisphere analyzer can be written as $\Delta E_{analyzer} \propto (\frac{w}{R_0} + \frac{a^2}{4})$, where $w$ is the width of the entrance slit, $a$ is the angular resolution and $R_0$ is the radius of the analyzer. The illustrative hemisphere has a central radius of 20 cm, while the need for magnetic shielding and vacuum chamber components results in a large total size. The mass of the analyzer alone can reach ~200 kg.

The sample stage is another critical component. ARPES sample stages often enable full 6-axes of motion for precise alignment and angular/*k*-space scanning. Temperature control from liquid helium cryogenic temperatures in the 4 K range, to a few tens or even hundreds of degrees above room temperature, is common. Certain experiments also benefit from the sample stage's ability to integrate electrical contacts for the application of fields and currents to samples. The cryogenic sample stage in Fig. 3 can be seen in the center of the lower cutaway view, holding a circular sample holder at the focus of the incoming light, approximately 35 mm in front of the spectrometer entrance. The stage itself and the large x-y-z manipulator mounted on top of the system provides six translational and rotational degrees of freedom, as shown.

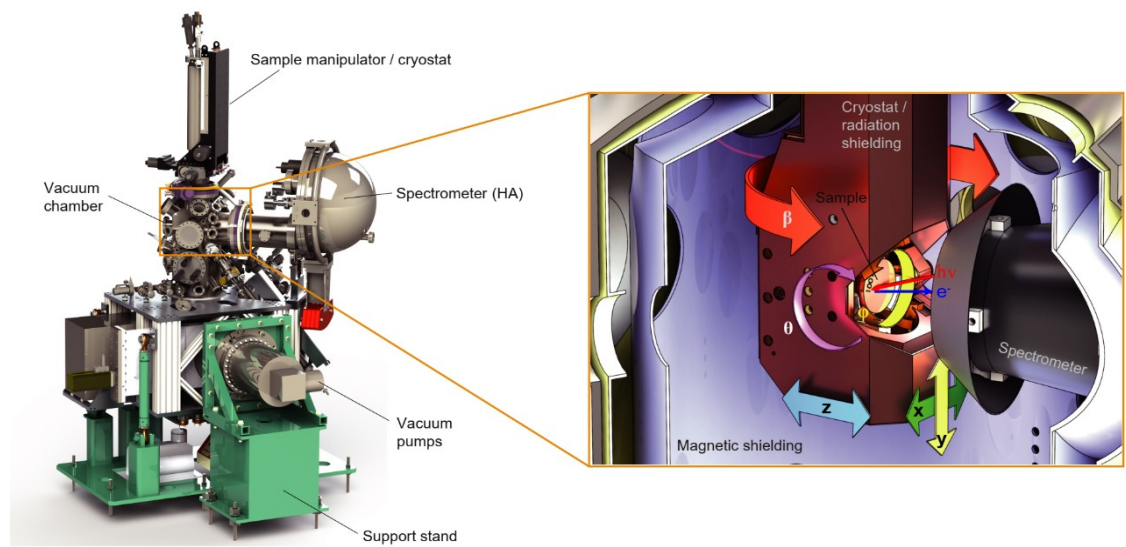

**Fig. 3. Schematic drawing of the ARPES endstation.** Left: a schematic diagram of the ARPES endstation. Right: the zoom-in image of the sample stage with six-axis translational and rotational degrees of freedom.

| Technique | | Advantages | Disadvantages | Experimental notes | Refs |
|---|---|---|---|---|---|
| High-resolution ARPES (HR-ARPES) | | Revealing low-energy physics; High momentum resolution up to 0.0008 Å$^{-1}$, energy resolution better than 1 meV | The use of VUV range photons (typically 7-60 eV) makes it extremely surface sensitive, while the lower energy photons limit the measurable momentum space | Typically utilizes light sources of 7-60 eV from synchrotron, laser sources with 6-7 or 11 eV, and Gas-discharge lamp sources. Usually performed at low temperatures to minimize thermal broadening near $E_F$, for example using $^4$He (~ 1.5 K) or $^3$He liquid (< 1 K) | 10-13,19 |
| Spin-resolved ARPES (SpinARPES) | Mott detector | Probing spin vector direction of electrons in both energy and momentum resolution, applicable to ferromagnetic materials or strong spin-orbit coupling (SOC) systems | Reduced efficiency, which requires lengthy data acquisition times | With the lower figure of merit ($F = 10^{-4}$-$10^{-5}$); The gold target survives for a long time | See Figure 4a and references 35-39 |
| | Very low energy electron diffraction (VLEED) detector | | | Higher figure of merit ($F = 10^{-2}$-$10^{-4}$); The surface of the magnetic target requires a sophisticated oxidation treatment and needs to be regenerated | |
| Nano- or micro-spot ARPES (NanoARPES/Micro ARPES) | | Probing small samples, materials with disorder or inhomogeneity, and phase-separated systems; Spatial resolution of micron scales and below (50-150 nm) | Requires specialized experimental set-ups; Reduced data acquisition due to the lower photon flux; limited space for sample movement or rotation | The light is tightly focused by the optics (e.g., a diffractive Fresnel zone plate); Piezo-driven stages and scanners are used with nm-scale precision | Figure 4b and references 25-29,31,94,95 |
| Time-resolved ARPES (TrARPES) | | Extending the measurements to the time domain, allowing to reveal of unoccupied states, dynamic properties on fs or ps time scales, and light-induced phenomena | Complicated experimental setups by merging ARPES and ultrafast pump-probe technique; Trade-off between energy and time resolution due to the energy-time uncertainty principle | Pump-probe configuration using two pulsed light pulses and a delay stage | Figure 4c and references 16-18,32-34,96-99 |
| Soft x-ray ARPES (SX-ARPES) | | Enabling detection of buried surfaces and interfaces; Covering a large $k_z$ range | Compromised energy and $k_{//}$ resolutions; Trade-off between resolution and bulk-sensitivity | $h\nu$ < 5-10 keV | 5,9,22,23,100 |
| Hard x-ray ARPES (HX-ARPES) | | | | $h\nu$ > 5-10 keV | 101 |

Table 2: Basic ARPES setup and variations of ARPES and methodologies.

**ARPES variations**

The ARPES technique has developed into various branches, including SpinARPES, Nano/MicroARPES, and TrARPES, which are capable of resolving spin, spatial, or time information, respectively. Fig. 4 gives an overview of these variations and their applications. The advantages and disadvantages of different variations of ARPES, including HR-ARPES, SX-ARPES, HX-ARPES, SpinARPES, NanoARPES/MicroARPES are summarized in Table 2. For more detailed information, see the supplementary information. With these variations, ARPES has the unique capability to directly reveal $E(k)$ with spin-, time-, and spatially-resolved information. By combining ARPES measurements with other techniques, complementary information can be obtained to reveal the materials' physical properties. For example, scanning tunnelling microscope (STM)[102] provides atomic-scale spatial resolution and allows detection of the unoccupied states with a much higher energy resolution, while being compatible with the application of magnetic field, gating, and extremely low temperatures. However, it lacks momentum resolution and is extremely surface sensitive. Resonant inelastic X-ray scattering (RIXS)[103] measures many-body excitations with both energy and momentum resolution, which can be applied to a large variety of samples including insulating samples, liquids, and polymers, however, it detects bulk properties with worse energy resolution compared with STM and ARPES. Ultrafast optical spectroscopy measures the transient reflectance or transmittance of materials with high time resolution. It is compatible with various experimental conditions such as ambient pressure. However, it detects the contributions from all electrons and lacks energy- and momentum-resolution.

**Multi-Modal ARPES**

The increasing branches of ARPES into new sensitivities and modalities has added to its range of applications in materials science. Its impact is further enhanced when multiple sensitivities can be simultaneously combined. For instance, simultaneous spin- and time-resolved ARPES experiments provide a unique probe of ultrafast spin dynamics for fundamental spin research[104,105], as well as information technology applications. Likewise, the prospect of spin-resolved NanoARPES could provide an important probe of spin physics in confined dimensions, for example, the topological edge states of quantum spin Hall systems. Combining ARPES with transport measurements allows direct correlation of the electronic structure with the resulting macroscopic material properties.

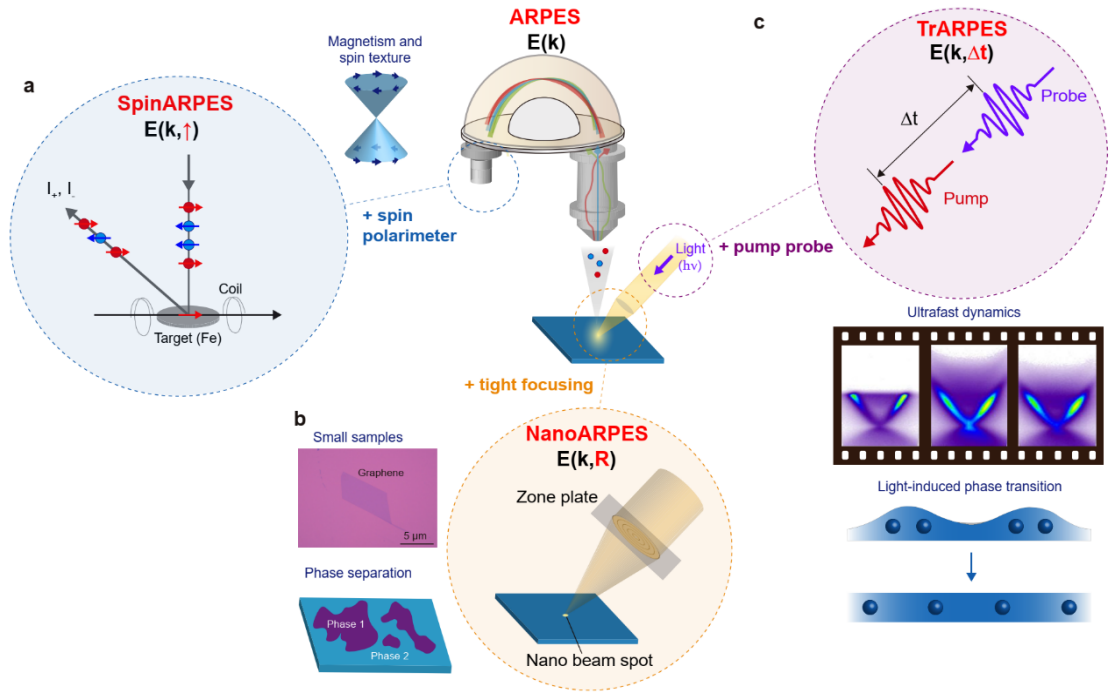

**Fig. 4. Basic principles and schematic drawing of ARPES variations.** The center panel shows the schematic drawing of ARPES. (a) SpinARPES can be performed by adding a spin polarimeter, which can detect the spin degree of freedom, an important characteristic for materials with magnetism and momentum-dependent spin textures. (b) NanoARPES can be performed by tightly focusing the photon beam to nanometer scale, for example with zone plate optics, which provides the spatial resolution to enable measuring of small samples such as exfoliated 2D materials and structures, and materials with intrinsic phase separation or inhomogeneity. (c) TrARPES can be performed with pulsed light sources by combing the pump-probe technique with ARPES. The ultrafast time resolution enables resolving transient dynamics, such as electron relaxation dynamics and the light-induced phase transition.

## Results

### Families of raw data

The raw data format of an ARPES experiment depends on the detector type. The most common is the integrated imaging detector, which delivers stacks of colored images — where different colors represent different photoelectron intensities — as a function of one or more scanned variables. Alternatively, the raw data can be recorded as a stream of individually measured photoelectron intensities. Each detection event is recorded as a row in a table of the resolved coordinates. Event-table raw data have a large volume and require a preprocessing step to access interpretable representations[106,107], but enable a-posteriori correction of hard-to-control experimental artifacts such as lens distortions[108], space charging[109,110] or source fluctuations.

**Fermi energy calibration**

In principle, the maximum $E_k$ of photoelectrons flying in the vacuum is the photon energy subtracted by the work function of the sample. The Fermi energy, $E_F$, is mostly determined by the work function of the photoelectron analyzer to which the sample is electrically connected. Therefore, the $E_F$ value of the ARPES data becomes identical for all materials grounded to the equipment as long as the same photon energy is used. This enables the use of reference material for determination of $E_F$ during data analysis. A commonly used material is evaporated polycrystalline gold. It is very stable in vacuum and has an almost constant density of states below $E_F$, resulting in an angle-integrated signal that can be fitted by the Fermi-Dirac distribution function. Consequently, the reference value of $E_F$ can be precisely determined by fitting the obtained spectrum to the Fermi-Dirac distribution function multiplied by a slope[1]. In practice, the extracted $E_F$ can vary slightly for each setup or day-by-day condition. Furthermore, the $E_F$ energy calibration can circumvent the uncertainty in the absolute value of the photon energy. In some cases, this uncertainty can be too large to satisfy the precision required for high-resolution ARPES measurements, for example at synchrotrons especially during photon energy scans. It is therefore important to obtain $E_F$ from a reference material before and after taking spectra for accurate data analysis.

**Energy, momentum and time resolution**

The energy resolution of an ARPES measurement can be estimated by fitting spectra from a reference material, usually evaporated gold, to the Fermi-Dirac distribution function at the measurement temperature, convoluted by the Gaussian function. The width of the obtained Gaussian function is the overall energy resolution $\Delta E$ of the ARPES system. The momentum resolution $\Delta k$ is not straightforward to obtain by fitting, because the momentum width is usually dominated by the intrinsic width of the sample. Instead, it is usually estimated from the angular resolution $\Delta\theta$ of the analyzer by $\frac{\Delta k}{k} = \frac{\Delta E}{E_k} + \frac{\cos\theta}{\sin\theta}\Delta\theta \approx \frac{\cos\theta}{\sin\theta}\Delta\theta$.

In pump-probe measurements, electrons are excited to the unoccupied states within the temporal envelope of the pump pulse and photoemitted within the temporal envelope of the probe pulses. The time resolution can be estimated by fitting time-trace data to a step function with exponential decay convoluted with a Gaussian function. The Gaussian function reflects the convolution of pump and probe pulses and its width defines the total time resolution. It is advisable to perform this analysis at the highest energy signal[111,112], rather than close to $E_F$, to exclude contributions from electron relaxation in the determination of the time resolution. The most reliable method is the observation of replica bands arising from the laser-assisted photoelectric effect[113,114] and the Floquet-Bloch states[99,115], allowing determination of both the pump-probe time overlap and the time resolution with high accuracy.

**Angle-to-momentum conversion**

The momentum information of electrons inside a solid can be deduced from the photoelectron's momentum outside the surface after photoemission. This is related to the measurement angle and experimental geometry in an ARPES experiment[1]. The angle-to-momentum conversion is important to obtain the momentum information.

From the experimental geometry shown in Fig. 5a, the momentum of photoelectrons outside the sample surface can be calculated by considering the rotation matrices[116,117] involving rotation of $\theta$ and $\beta$ angle, $M_\theta$ and $M_\beta$ as:

$$\begin{pmatrix} k_{x,out} \\ k_{y,out} \\ k_{z,out} \end{pmatrix} = kM_\theta M_\beta \begin{pmatrix} \sin\alpha \\ 0 \\ \cos\alpha \end{pmatrix} = k\begin{pmatrix} \cos\beta & 0 & \sin\beta \\ 0 & 1 & 0 \\ -\sin\beta & 0 & \cos\beta \end{pmatrix}\begin{pmatrix} 1 & 0 & 0 \\ 0 & \cos\theta & \sin\theta \\ 0 & -\sin\theta & \cos\theta \end{pmatrix}\begin{pmatrix} \sin\alpha \\ 0 \\ \cos\alpha \end{pmatrix} =$$

$$k\begin{pmatrix} \cos\beta\sin\alpha + \sin\beta\cos\theta\cos\alpha \\ \sin\theta\cos\alpha \\ -\sin\beta\sin\alpha + \cos\beta\cos\theta\cos\alpha \end{pmatrix} \text{ (eq. 4)}$$

where $k = \frac{1}{\hbar}\sqrt{2m_e E_k} = 0.512\sqrt{E_k}$. Here the unit of $k$ and $E_k$ are Å$^{-1}$ and eV respectively, and 0.512 is the numerical value of $\sqrt{2m_e}/\hbar$.

The in-plane momentum of electrons inside the material, $k_{x,in}$ and $k_{y,in}$, can then be obtained directly from the in-plane momentum conservation. This is because the in-plane momenta are good quantum numbers due to in-plane translational symmetry of the crystal. The out-of-plane momentum is, however, not conserved during the photoemission process due to the lack of translational symmetry along the sample surface's normal direction. To extract the out-of-plane momentum $k_{z,out}$, a free-electron approximation is assumed for the final state of photoexcitation[118], with a potential difference $V_0$ for electrons outside and inside the sample. For example, $\frac{\hbar^2 k_{z,out}^2}{2m_e} + V_0 = \frac{\hbar^2 k_{z,in}^2}{2m_e}$, where $V_0$ is a fitting parameter called the inner potential[54,118]. The momentum of electrons inside the sample is calculated by:

$$\begin{cases} k_{x,in} = k_{x,out} = 0.512\sqrt{E_k}(\cos\beta\sin\alpha + \sin\beta\cos\theta\cos\alpha) \\ k_{y,in} = k_{y,out} = 0.512\sqrt{E_k}\sin\theta\cos\alpha \\ k_{z,in} = \sqrt{k_{z,out}^2 + 0.512^2 V_0} \end{cases} \text{ (eq. 5)}$$

A different $k_z$ can be obtained by changing the photon energy. For 3D states, the $k_z$ dispersion is significant, while 2D electronic states have a constant intensity and energy position at all photon energies. Empirically, the inner potential $V_0$ is between a few eV to 20 eV, usually determined by aligning the high symmetry points in the $k_z$ map or dispersion.

The conversion in (eq. 5) is quite complicated, involving all three angles. For small $\theta$ angle, $cos\theta \approx 1$, and $k_{x,in}$ can be approximated by $k_{x,in} \approx 0.512\sqrt{E_k}(cos\beta sin\alpha + sin\beta cos\alpha) = 0.512\sqrt{E_k}sin(\alpha + \beta)$. Namely, $\alpha$ and $\beta$ are equivalent, and the angle-to-$k$ conversion is greatly simplified. However, such approximation works only for small $\theta$ angle. For large $\theta$ angle, the accurate conversion using (eq. 5) has to be used[117], see illustrated example in a Kekulé-ordered graphene[92] in Fig. 5b.

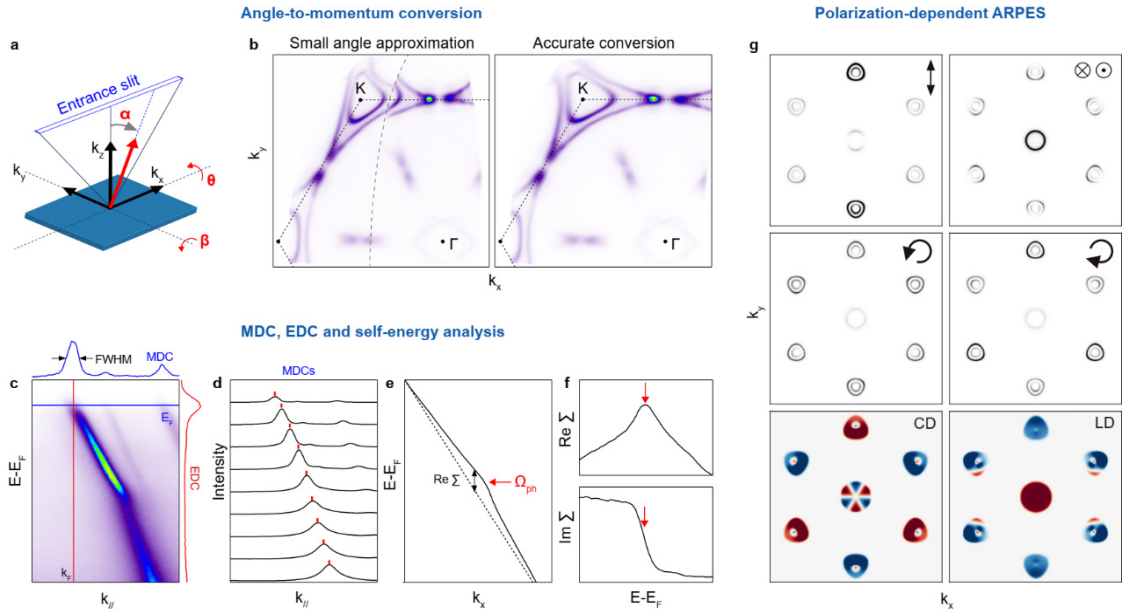

**Fig. 5. ARPES data analysis**. (a) Configuration of sample rotation with both large polar ($\theta$) and tilt angle ($\beta$). (b) Converted FS maps by approximation at small rotation angles and correction at large rotation angles, respectively. Each FS is combined with two maps acquired at different tilt angles, and the gray broken curve indicates the connection line between the two maps. (c), ARPES measured 2D intensity map, with the extracted MDC and EDC shown by blue and red curves. (d), MDC stacks with fitting results appended by red marks to extract the dispersion shown in (e). (f), Extracted real and imaginary parts of Σ. Red arrows indicate the phonon energy. (g) Isoenergy cuts at normal emission of the valence band of $WSe_2$ at -0.7 eV with in-plane vertical and out-of-plane polarization, respectively. The middle two panels are the same as the upper panels but with circularly polarized light of opposite helicities. The bottom panels show the CDAD and LDAD images. All calculations are performed using the chinook software package[119].

**Data analysis**

In real materials, photoemission is a many-body process due to electron interactions in solids. It can be described by a rigorous approach called the one-step model[120] where the bulk, surface, and vacuum are included in the Hamiltonian. Such a one-step model requires advanced computational methods and is difficult to grasp intuitively. Another approach, called the three-step model, is often used to intuitively describe the photoemission process[1,118]. It has three independent steps: optical excitation of the electrons inside the solid; travel of the excited electron to the sample surface; escape of the photoelectrons into the vacuum.

The measured ARPES intensity is given by the product of these three corresponding steps: the probability of optical excitation, the scattering probability of photoelectrons when traveling to the sample surface, and the probability of photoelectrons to escape into the vacuum. The first step includes all information about the excitation process and is the most important for the physical interpretation of the data. The second step describes the scattering of photoelectrons when approaching the sample surface. This is related to, and can be approximated by, the effective mean free path, as shown in Fig. 1d.

**Dispersion E(k) from EDCs and MDCs analysis**

ARPES results typically show a 2D intensity map as a function of $E_B$ and $\boldsymbol{k}$, Fig. 5c. In the sudden approximation[1,121], the photoemission process in the first step is assumed to occur without any interaction between the photoelectron and the system left behind. This assumption is more accurate for photoelectrons with higher kinetic energy. The measured ARPES intensity can be written as $I(\boldsymbol{k},\omega) = I_0(\boldsymbol{k},v,\boldsymbol{A})A(\boldsymbol{k},\omega)f(\omega)$, where $\boldsymbol{k}$ is the electron momentum, $\omega$ is the electron energy with respect to the Fermi energy, $f(\omega)$ is the Fermi-Dirac function, $I_0(\boldsymbol{k},v,\boldsymbol{A})$ is proportional to the squared one-electron matrix element $\left|M_{k_f,k}\right|^2$ and $A(\boldsymbol{k},\omega)$ is the single-particle spectral function which contains information about the dispersion and photohole lifetime.

There are two methods to extract the dispersion from the ARPES intensity map: energy distribution curve (EDC) analysis and momentum distribution curve (MDC) analysis[122]. The EDC shows ARPES intensity as a function of energy at fixed momentum (red curve in Fig. 5c), which gives an overview of the spectral function and the line shape. The EDCs are effectively photoemission spectra for precise $k_x$, $k_y$ values. They have a clear physical interpretation and are used to investigate many details of the electronic structure. For example, they allow observation of the presence or absence of a well-defined sharp quasiparticle peak, and extraction of the energy gap induced by the superconducting (SC) or charge density wave (CDW) transition. For other applications, such as the precise determination of band dispersion by peak fitting, the complex information incorporated in an EDC becomes a drawback. EDCs are affected by the Fermi-Dirac distribution near the Fermi energy, and there is an energy-dependent background — the Shirley background[123] — which is not included in the spectral function. This complicates the fitting functions and, when possible, a different approach based on MDC is used.

The MDC shows ARPES intensity as a function of momentum at fixed energy (blue curve in Fig. 5c). MDC usually has a more symmetric line shape and can be fitted by a Lorentzian function, allowing extraction of both the peak position and full width at half maximum. This gives information about the scattering rate Γ, which is inverse to the quasiparticle lifetime[124]. By fitting the MDC stacks (Fig. 5d), the peak position can be extracted at each energy to obtain the dispersion (Fig. 5e).

For SpinARPES measurements, data analysis typically involves taking the differential signal of EDCs or MDCs measured by different spin channels. Similarly, in TrARPES, differential EDCs or MDCs between positive and negative delays are studied to individuate dynamical changes in the electronic structure. For more detailed information, see sections 2.1 and 2.2 in the supplementary information.

**Important band parameters and self-energy analysis**

The obtained electronic structure $E(k_x, k_y, k_z)$ enables extraction of important band parameters introduced in solid-state physics, such as the Fermi velocity, effective mass, bandwidth, bandgap and more.

Analysis of ARPES data also provides information on the many-body interaction in solids. The electrons or holes propagating a many-body system are correlated with each other and coupled to various collective modes[89,125,126] such as phonons, magnons, and plasmons. For example, many-body interactions including electron-electron[127], electron-phonon[128] and electron-plasmon coupling[89] are reported to induce different band renormalization in monolayer graphene. These complex effects are all expressed as a single function called self-energy $\Sigma(\mathbf{k}, \omega)$, which is contained in the single-particle spectral function by:

$$A(\mathbf{k}, \omega) = \frac{1}{\pi} \frac{|\mathrm{Im}\,\Sigma(\mathbf{k},\omega)|}{[\omega - \varepsilon_k - \mathrm{Re}\,\Sigma(\mathbf{k},\omega)]^2 + [\mathrm{Im}\,\Sigma(\mathbf{k},\omega)]^2} \quad \text{(eq. 6)}$$

Here the real and imaginary parts of the self-energy, ReΣ and ImΣ, reflect the energy renormalization and lifetime of electrons due to the many-body interaction, respectively[129], and they are related to each other by the Kramers-Kronig relation[130]. Since self-energy is fundamental for understanding the electronic properties of solids, its determination is one of the main subjects of ARPES study.

The MDC fitting analysis is now commonly employed to extract the self-energy from ARPES data[122,130]. The energy dispersion of bare electrons free from interaction could be approximated as $\varepsilon_k \approx v_F^0(\boldsymbol{k} - k_F)$ for $\boldsymbol{k}$ near $k_F$ with the bare Fermi velocity $v_F^0$, leading to Re $\Sigma(\omega) \approx \omega - v_F^0[k_m(\omega) - k_F]$. Here, Re $\Sigma(\omega)$ represents the energy difference between the renormalized band and the bare band (see Fig. 5f), which energetically narrows the bare band and yields a kink structure at the mode-coupling energy of bosons such as the phonons and magnons (indicated by the red arrow) when the interaction is strong enough. The determination of $v_F^0$ is challenging, but possible by self-consistently tuning the parameters of Re$\Sigma$ and Im$\Sigma$ via the Kramers-Kronig relation to eventually reproduce the ARPES intensity map[129]. On the other hand, Im$\Sigma(\omega) \approx -v_F^0 \Delta k/2$ can be extracted by spectral width — $\Delta k$: full width at half maximum of MDC peak — which relates to the scattering rate $\Gamma \approx 2\text{Im}\Sigma$ (or lifetime $\tau = \hbar/\Gamma$), and broadens the spectrum especially at high binding energies due to a shorter photohole lifetime[122,131].

For TrARPES measurements, the carrier dynamics[111,132-134], band renormalization[96,97,99,135,136] and self-energy[137,138] at different time delays can be extracted. Moreover, the dynamical orbital tomography[139] can also be obtained. Similarly, spin-dependent self-energy parameters can be extracted from spinARPES data. For more detailed information, see section 2.1 and 2.2 in the supplementary information.

**Dipole Matrix element effect**

Besides $A(\boldsymbol{k}, \omega)$, the ARPES signal is also modulated by the dipole matrix element by $I_0(\boldsymbol{k}, \nu, \boldsymbol{A}) \propto \left|M_{k_f,k}\right|^2$. The matrix element is the expectation value of the dipole operator between the initial state $|\Psi_{n,k}\rangle$ and the final state $|\Psi_{nf,kf}\rangle$, $M_{k_f,k} = -i\frac{\hbar e}{mc}\langle\Psi_{n_f,k_f}|e^{ik_{hv}\cdot r}\hat{\varepsilon}\cdot\nabla|\Psi_{n,k}\rangle$, where $\hat{\varepsilon}$ is the light polarization.

Three main factors influence $M_{k_f,k}$: the wavefunction properties, the photon energy, and its polarization or direction of incidence. Firstly, the orbital symmetry and texture of the wavefunction lead to intensity modulations within the Brillouin zone (BZ), while the crystal space group symmetry leads to signal suppression or enhancement between equivalent points in different BZs[140]. In a simplified picture, the intrinsic part of $M_{k_f,k}$ is conceivable as the Fourier transform of the localized Wannier function associated with the initial Bloch state $|\Psi_{n,k}\rangle$. The Wannier function can be approximated as hybridized atomic orbitals in a tight binding framework[55]. The ARPES intensity can then be viewed as the result of the coherent superposition of photoelectron waves carrying spatial and phase information of the original hybridized orbital, modulating the photoemission intensity in the 3D $k$-space. Dedicated tutorials on this topic are provided in these articles[55,119]. Secondly, changes in the photon energy correspond to ARPES spectra at different values of $k_z$ in the 3D BZ, which arise from the orbital and crystal symmetries.

Finally, the light polarization is intrinsically coupled to the experimental geometry. This can provide useful information for orbital tomography, in particular for experimental apparatus that does not require sample movement to map the BZ. Examples include the wavefunction reconstruction of molecular systems on surfaces or in crystals[139,141], the exploration of orbital character in complex crystals[142-144], and the study of band topology in quantum materials[67,145]. The dominant term in $M_{k_f,k}$ is the projection of $\varepsilon$ along the photoelectron propagation direction. The intensity variations arising from changes in light polarization in fixed experimental conditions generate a family of observables called dichroism in the angular distribution (DAD). Figure 5g shows the calculations with linear and circularly polarized light for the idealized case of normal incidence and normal emission. The linear DAD (LDAD) illustrated in the top panels of Fig. 5g show the asymmetry between ARPES spectra measured with s- and p-polarized light that can be related to the directional character of the electronic states[146-149]. The circular DAD (CDAD) shown in the middle panels is sensitive to the local handedness of the wavefunction and has been linked with the local Berry curvature [146,147,150,151] and spin-orbital texture[67]. Caution is needed as the experimental geometry itself can introduce a handedness and result in CDAD signals[55]. By detecting the ARPES intensity changes under symmetry operations on the crystal, it is possible to retrieve information on the local orbital texture, such as the orbital pseudospin obtainable by time-reversal DAD[144,152].

Matrix elements can be convolved with many-body interactions[153], entirely suppressing the intensity of individual bands[154], or disturbing the dynamical evaluation of observables[155]. In set-ups requiring continuous sample movement to map the BZ, the changing experimental geometry makes the behavior of $M_{k_f,k}$ highly counterintuitive. The matrix element effect in ARPES is sensitive to the photon energy[156] and polarizations[146-149]. Therefore, techniques to mitigate the intensity suppression have been developed — for example, enlarging the dataset with different photon energies[157], different polarizations[158,159], and higher-order BZs — which can help to pinpoint the matrix element origin of a feature suppression or enhancement. For orbital selective measurements[160,161], the light polarization (s- or p-polarization) and the electron orbitals of interest need to have the same parity with respect to a virtual mirror plane set for the ARPES system. This requires careful alignment of the high-symmetry directions during the ARPES measurements.

**Statistical analysis and errors**

Since ARPES measures the number of photoelectrons $N$ at specific energy and $k$ that obeys Poisson distribution, it is important to consider the statistical error bar of the photoelectron counts ($\sqrt{N}$) while extracting physical parameters from the obtained ARPES data. For proper analysis of the ARPES data, other experimental uncertainties, such as energy resolution, $k$ resolution, stability of electron $E_k$, and photon energy during the time of ARPES measurement, need to be taken into account. Owing to the surface-sensitive nature of ARPES, degradation of the sample surface can sometimes cause energy shifts and broadening of peaks. As a result, surface degradation needs to be carefully checked when accumulating statistics in long scans. All these issues are reflected as total error bars of the ARPES data. Total errors need to be carefully considered, especially when quantitatively discussing key band parameters whose energy or $k$ scale is close to experimental uncertainties. This is often the case for extracting energy gap associated with SC pairing, CDW, magnetic gap opening, and magnitude of band splitting due to the interlayer coupling, exchange coupling, spin-orbit coupling (SOC) and more. To reliably estimate the band parameters without relying on human determination, statistical methods such as Bayesian inference[162] can be also adopted for the data analysis.

**Applications**

Over the past few decades, ARPES measurements have been extensively applied to a wide range of solid-state crystals, leading to major progress in the understanding of the electronic, spin structure as well as the ultrafast dynamics in the nonequilibrium state. This section discusses examples of three families of materials whose physics has been significantly advanced by ARPES measurements, including superconductors, topological materials, two-dimensional materials and heterostructures.

**Superconductors**

ARPES is one of the leading experimental techniques to study the physics of high-temperature superconductors[1,2,24]. Stimulated by the discovery of cuprate superconductor $La_{2-x}Ba_xCuO_4$ in 1986[163], ARPES achieved a drastic improvement in energy resolution in the 1990's. This leads to the observation of key electronic states essential for understanding the nature of superconductivity, such as the FS[164], and a large superconducting gap showing strong anisotropy (left panel of Fig. 6a)[74,165] and a pseudogap above $T_c$ [166-168]. Perhaps the most intensively studied bulk crystalline solid by ARPES is the bismuth-based high-temperature copper oxide (cuprate) superconductor, $Bi_2Sr_2Ca_{n-1}Cu_nO_{2n+4+\delta}$ ($n$ = 1-3), known as BSCCO. It has a stable, clean surface which is easily obtained by *in situ* cleaving of a crystal due to weak van der Waals bonding. There is also fundamental scientific interest in the mechanism of high-temperature superconductivity and a variety of exotic quantum phases surrounding superconductivity.

To derive the mechanism of superconductivity, it is essential to experimentally determine superconducting pairing symmetry because it reflects the types of media that promote Cooper pairing, such as phonons, magnons, and charge fluctuations. ARPES has played a pivotal role in revealing the superconducting pairing symmetry of cuprates by directly observing the $k$-dependence of the superconducting gap associated with the pairing. Earlier ARPES studies of $Bi_2Sr_2CaCu_2O_{8+\delta}$ (Bi2212) clarified $d_{x^2-y^2}$-like anisotropic superconducting gap with a gap maximum around the $(\pi, 0)$ point of $CuO_2$ BZ, called antinode, and the gap node along the $(0, 0)$-$(\pi, \pi)$ line, diagonal to the Cu-O bond[74,165], whose detailed $k$-dependence is summarized in the right panel of Fig. 6a[169]. Gaps for both bonding and antibonding bands associated with interlayer coupling are clearly resolved in Fig. 6a. This suggests spin-mediated pairing, unlike conventional Bardeen-Cooper-Schrieffer (BCS) superconductors[170] which show fully gapped $s$-wave symmetry due to the phonon-mediated pairing, though this is still under debate. In addition to the $d$-wave superconducting gap, ARPES has also observed an energy gap above $T_c$, called a pseudogap. In the underdoped region of the electronic phase diagram, the pseudogap was observed as anomalies in various physical quantities such as density of states, magnetic, transport, and thermal properties. Its origin has long been at center of debate. Recent improvement in the energy and $k$ resolution enabled by the development of high-flux light sources, such as the quasi-continuous-wave (QCW) vacuum ultraviolet (VUV) laser, has provided new opportunities to investigate the gap structure in great detail[171]. By obtaining EDCs at various temperatures with the aid of precise temperature control, it is possible to determine the gap closing temperature at different $k_F$ points on the FS. The results indicate that the $d_{x^2-y^2}$-like energy gap with the point node persists even above $T_c$, suggesting the existence of pairing-induced pseudogap, although a pseudogap of different origin may also exist at the antinode[172-175]. ARPES also disclosed key many-body interactions observable as a sudden bending, or kinks, in the band dispersion near $E_F$ [125,176] (Fig. 6b). This dispersion kink is unpredicted from single-particle approximations such as density-functional-theory (DFT) calculations commonly observed in different materials classes. It originates from the coupling of electrons to collective modes such as phonons and magnons[1,177], which are important interactions related to the mechanism of superconductivity.

SpinARPES was recently applied to Bi2212, to reveal the $k$-dependent spin texture that circles the BZ center[178]. This suggests a possible role of the unconsidered SOC in superconductivity. The role of SOC needs to be investigated further both experimentally and theoretically. The rapid development of TrARPES over the past decade has also extended the investigation of superconductors to the dynamics of Cooper pairs in the nonequilibrium state. By perturbing the superconducting state using ultrashort femtosecond laser pump pulse, quasiparticle spectral weight and superconducting gap structures were studied as a function of delay time[135,179,180]. For example, in Bi2212, the superconducting gap along the off-nodal cut exhibits a dynamical change upon photo-excitation[33,135,179] (Fig. 6c), signifying a partial closure of the superconducting gap accompanied by a reduction of quasiparticle weight. By systematically investigating the gap size as a function of delay time at different $k_F$ points, an interesting difference in the quasiparticle recombination rate between nodal and off-nodal regions was observed[135,179].

Besides cuprates, ARPES has been widely applied to many other superconductors with a smaller superconducting gap, such as iron-based superconductors[181-183], low-$T_c$ superconductors, and topological superconductor candidates. Many key features in both bulk crystals and thin films of iron-based superconductors were identified by ARPES. These are represented by the observation of Fermi-surface-dependent superconducting gap (Fig. 6d)[184-188]; reconstruction of band structure associated with electronic nematicity[189,190]; topological Dirac-cone surface states hosting a pairing gap due to the superconducting proximity effect[191,192]; and a single large electron pocket hosting an $s$-wave superconducting gap[193-196] and replica band associated with interfacial electron-phonon coupling in a single-layer FeSe-based film[197-199]. TrARPES has been also applied to single-layer FeSe, where a time-dependent oscillation of the band energies in Fig. 6e was observed, which is taken as a signature of electron-phonon coupling enhanced by the electron correlation[200].

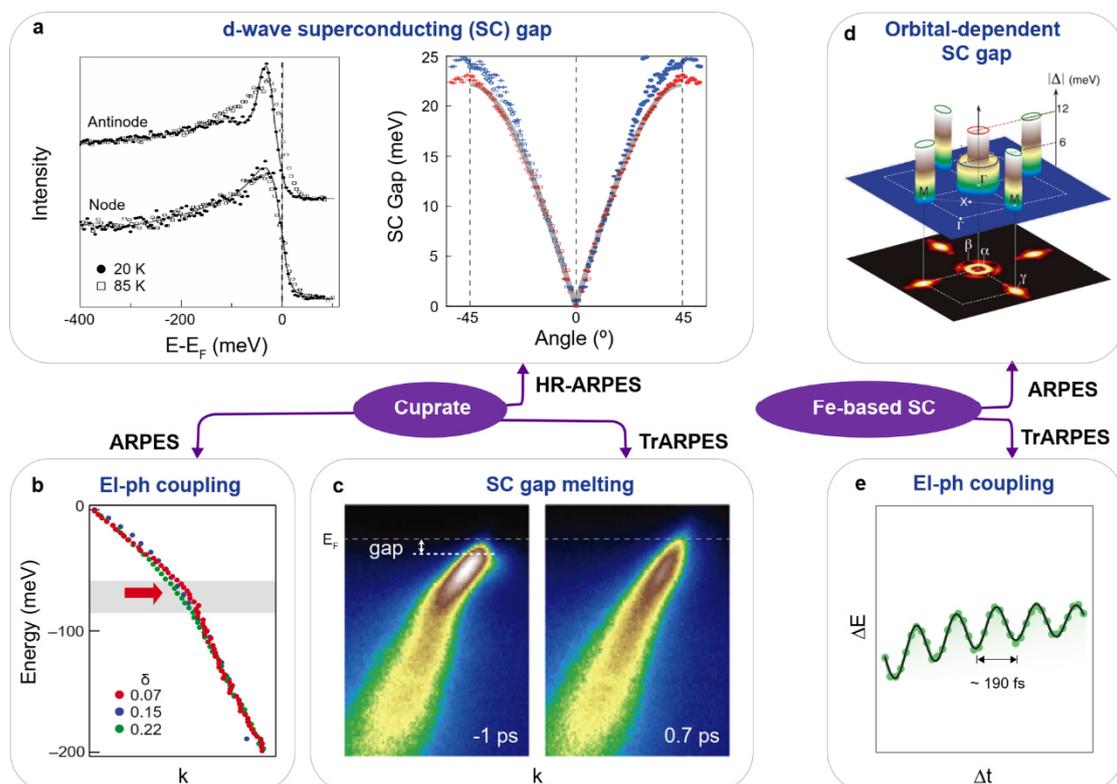

**Fig. 6. ARPES examples on cuprates and iron-based superconductors.** (a) EDCs at the antinode and node in Bi2212[74] and the $d_{x^2-y^2}$-wave superconducting gap of Bi2212 for bonding and antibonding bands[169]. (b), Electron-phonon coupling induced kink in the band dispersion[125]. (c) Melting of the gap in the superconducting state upon photo-excitation of Bi2212[33]. (d) The fermi-surface-dependent superconducting gap in $Ba_{0.6}K_{0.4}Fe_2As_2$[184]. (e) Time-dependent oscillation of band energies associated with the strong electron-phonon coupling in FeSe[200].

**Topological materials**

The electronic properties of topological materials, categorized by their bulk band topology, is an area of great interest in materials science[43,201-203]. This research field has been flourishing following theoretical predictions and subsequent experimental realization of topological insulators (TIs)[42]. ARPES played an important role in experimental exploration of TIs[204-206] and leads the field in revealing the nature of the TI state and other topological states, such as topological crystalline insulators[207-209], Dirac semimetals[210-212], Weyl semimetals[45,213-216], topological nodal-line semimetals[217,218], and topological superconductors[191,219].

While TIs have a bulk bandgap similar to a normal insulator, they form metallic states on their edges or surfaces due to a strong spin-orbit interaction, leading to the inversion of the conduction and valence bands. ARPES with surface sensitivity is a powerful experimental technique to directly observe the topological surface state. Figure 7 shows some examples of ARPES results on topological materials. $Bi_2Se_3$ is one of the most famous and well-studied compounds, which exhibits a single Dirac dispersion at the zone center[204,205] as shown in Fig. 7a[205]. This compound is categorized as a strong TI with a topological index of $Z_2 = 1$[220-222].

SpinARPES measurements have been performed in TIs such as $Bi_2Se_3$ and $Bi_2Te_3$, revealing the helical spin texture[223-226]. In addition, TrARPES was used to study photo-induced dynamical electronic states[111,227] and light-induced emerging phenomena. The most remarkable example would be the observation of Floquet-Bloch states[99,115], where the electronic bands strongly couple with photons and become periodic in momentum and energy. Floquet states can be viewed as a time analog of the Bloch states, and they were experimentally realized for the topological Dirac band of $Bi_2Se_3$ as displayed in Fig. 7b[99,115]. In addition to the Floquet-Bloch bands periodic both in momentum and energy, hybridization gaps are observed at the band crossing points. Moreover, the time-reversal symmetry can be broken by pumping with circularly polarized light, resulting in an opening of the gap at the Dirac point[99,115]. By combining TrARPES with SpinARPES, the detailed spin-polarization of the unoccupied topological surface state can also be revealed (Fig. 7c)[104].

In contrast to many ARPES reports on strong TIs, there are fewer studies of the weak TI [228-230]. The main difficulty in identifying a weak TI is that the topological surface state emerges only on a particular surface — usually the side crystal surface not naturally cleavable — making it difficult to observe by ARPES. β-$Bi_4I_4$ is a rare case where both top and side crystal surfaces are naturally cleavable due to the structure built from 1D chains[228,231]. Therefore, the top and side surfaces can be both exposed for ARPES measurements. In such a case, NanoARPES with spatial resolution can resolve the topological surface states that only appear on the side cleaving surface (Fig. 7d)[228].

Another important topological material system that has been advanced by ARPES is Weyl semimetals[44,45,214-216]. The Weyl semimetal states require breaking at least one of the time-reversal and spatial-inversion symmetries. The latter was first discovered in a non-centrosymmetric material, TaAs. TaAs has exotic Weyl nodes[213,214], characteristic topologically protected Fermi arc surface states[214-216] (Fig. 7e, left panel) and intriguing spin-polarization[232] (Fig. 7e, right panel), which were revealed by SX-ARPES, ARPES, and SpinARPES.

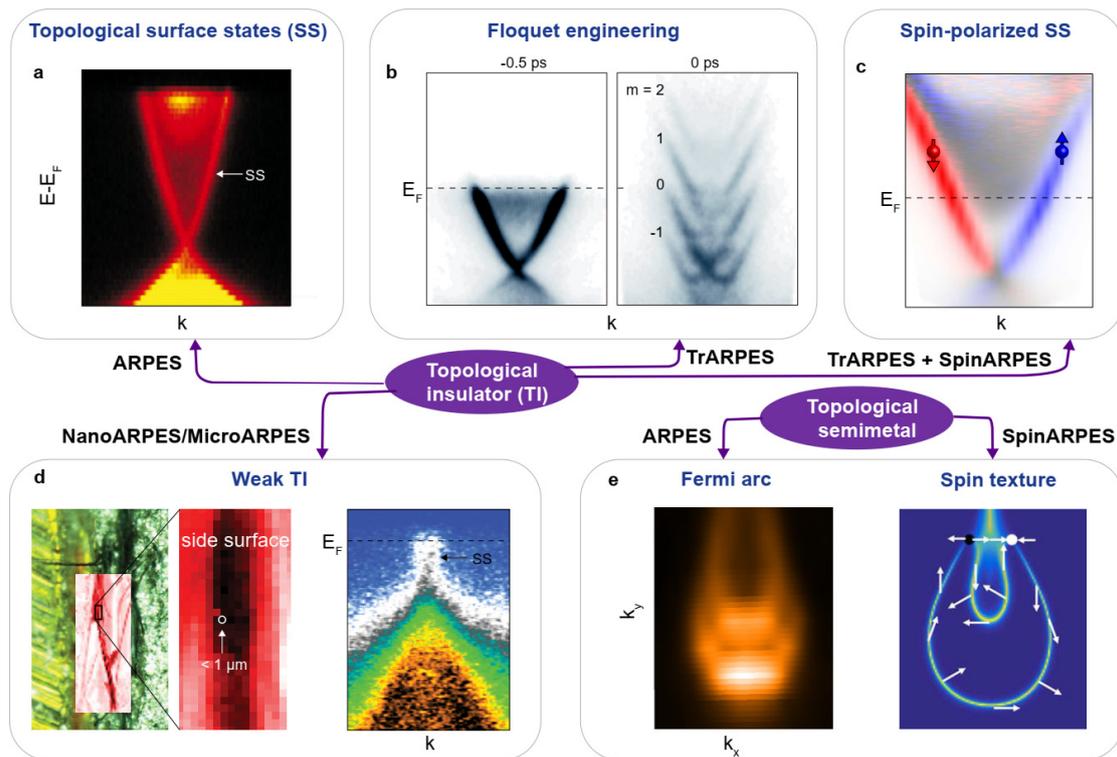

**Fig. 7. ARPES results on topological materials.** (a) ARPES measured dispersion image of $Bi_2Se_3$ to show the topological surface state[205]. (b) Bloch-Floquet states observed on $Bi_2Se_3$[115]. (c) Spin polarization observed in the topological surface state of $Bi_2Se_3$[104]. (d) The optical image, spatial intensity maps, and dispersion image measured on a weak topological insulator β-$Bi_4I_4$ by Nano-ARPES[228]. (e) The ARPES measured FS plot of the Weyl semimetal TaAs[215], and illustration of the spin texture of the Fermi arc[232].

## 2D materials and heterostructures

Dimensionality is one of the most important parameters that determines the material properties. 2D materials with unique layered structure and weak interlayer van der Waals interactions have gained interests over the past two decades. They exhibit interesting physics that is rare or non-existent in 3D systems. The first successfully exfoliated 2D material was graphene[233,234], which has a unique electronic structure described as massless Dirac fermions[235,236]. With new advances in synthesis, other 2D systems — including transition metal dichalcogenides (TMDCs) and 2D heterostructures — have been discovered[51,52,237].

ARPES has been established as an insightful technique that can be used in, for example, the direct observation of Dirac cones in graphene[84,238], or the evolution of electronic structure with number of layers[239,240]. Further exploration of the linear dispersion has revealed unexpected electronic correlations arising from coupling with bosons[89] and the surprising complexity of substrate interactions which can induce a gap opening at the Dirac point[238]. By performing TrARPES measurement, the hot carrier dynamics that contain details of the relaxation mechanisms[241] and the dynamical electron-phonon coupling[137,242] have been revealed.

ARPES has also played an important role in studying other 2D systems, such as the TMDCs, including the layer-dependent band structure[243,244], spin-vally[245-247] and spin-layer locking[248], and light-induced phase transiton[249,250]. For example, an indirect-to-direct band-gap transition has been observed in $MoS_2$ when going from bulk to monolayers[243,244] (Fig. 8a). Owing to the high excitonic $E_B$ arising from 2D confinement, TMDCs have offered a platform for studies of exciton physics by combining TrARPES with MicroARPES, with the direct observation of the excitons[251-254] (Fig. 8b, left panel) and the determination of key features of the excitonic wavefunction[255,256]. The most frequently observed emergent order in 2D systems is the formation of CDWs, where ARPES has been used to demonstrate the CDW folding of the bands, the formation of gaps in the FS, and the formation of shadow bands[257,258]. The use of TrARPES has enabled tracking of occupied and unoccupied states during the melting phase (Fig. 8b, right two panels), alongside tracking the order parameter evolution and observation of coherent oscillations, an excitation known as the amplitude mode[97,259,260]. Time- and angle-resolved two-photon photoemission (2PPE) has also been used to reveal the image-potential states and molecular states of 2D materials, providing another perspective of the ultrafast dynamics in low-dimensional systems[261].

Many 2D materials and heterostructures, especially modern field-effect devices, are obtained by mechanical exfoliation methods, with a limited sample size of a few to tens of micrometers. The technical progress of NanoARPES/MicroARPES has brought an ability to directly observe the electronic states of small samples[86,262]. Electrostatic gating can be applied to overcome a major limitation of ARPES, that it probes only the occupied electronic states, while also providing a powerful means of investigating phenomena and band structure engineering. For example, the band dispersion with a shift of the Fermi level across the Dirac point under gate voltages can be revealed, Fig. 8c[86,262].

The weak interlayer coupling by van der Waals forces in 2D materials enables exfoliation and artificial re-stacking of atomic-layered 2D materials to form various kinds of van der Waals heterostructures[51]. This introduces the potential to realize new physical phenomena and for future device applications. This research field is rapidly increasing in maturity, with new findings in moiré physics and twistronics[263-265]. NanoARPES/MicroARPES with a focused beam can directly reveal the band structure engineering, which is fundamental to the physics of heterostructures. For example, the Moiré potential in graphene/h-BN leads to second-generation Dirac cones (SDCs)[266] and gap openings. Mirrored Dirac cones[267] emerge in a new quasicrystal, formed by 30°-twisted bilayer graphene, analogous to quasicrystals with 12-fold symmetry[268] (Fig. 8d, left two panels). For magic-angle twisted bilayer graphene (MABLG) which shows superconductivity[269] and Mott insulating state[270], NanoARPES measurements reveal a flat band near $E_F$[271,272], which is essential to the physics of magic-angle twisted bilayer graphene (Fig. 8d, right panel).

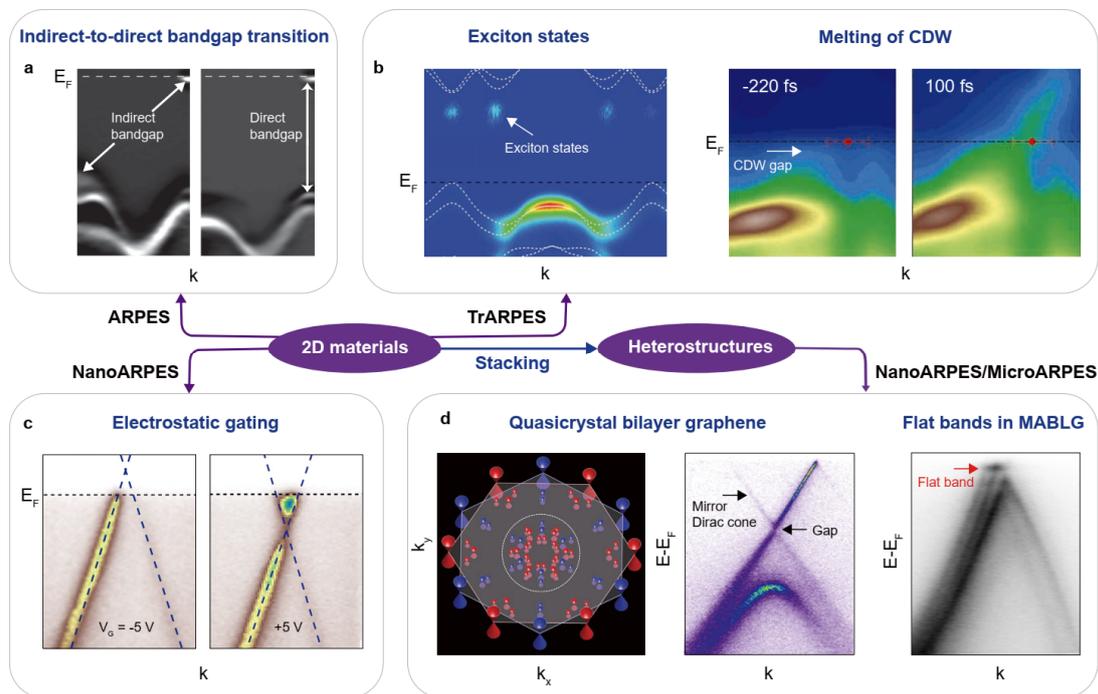

**Fig. 8. 2D materials and heterostructures.** (a) Indirect to direct bandgap transition in $MoSe_2$[244]. (b) Observation of direct and indirect excitonic states in exfoliated monolayer $WSe_2$ with the exciton at a 0.5 ps delay[251]. The right two panels show the melting of the CDW gap in $TbTe_3$ with 800 nm optical excitation[97]. (c) Electrostatic gating of graphene measured by

MicroARPES[86]. (d) The 12-fold rotational symmetry and band hybridization and in 30°-twisted quasicrystal bilayer graphene[267,268]. The right panel shows the flat band observed in magic-angle twisted bilayer graphene[272].

**Buried surface or interface**

While VUV-ARPES is surface-sensitive because of the short probing depth (Fig. 1d), SX-ARPES, with enhanced probing depth, can access systems buried a few atomic layers behind the top surface, such as the heterointerface and surface-capped samples. With the ability to resonantly probe valence *d*- and *f*-states, SX-ARPES can provide important information for strongly correlated systems such as Mott insulators, superconductors, and magnetic systems. SX-ARPES has been successfully applied to the interface between two band insulators, $LaAlO_3$ (LAO) and $SrTiO_3$ (STO)[273,274], a famous example of an interfacial system where the superconducting mechanism was intensively debated. The superconductivity is associated with the mobile 2D electron gas (2DEG) at the buried interface, which can hardly be accessed by VUV-ARPES. Enhancing the cross-section of Ti 3*d* orbital by tuning the photon energy to the 2*p*-3*d* resonance condition, SX-ARPES has clarified metallic Fermi edge cut-off together with the Fermi-surface topology associated with the interfacial 2DEG states. This resulted in identification of the 2DEG origin from the one-unit-cell-thick STO layer at the interface[275,276]. SX-ARPES was also applied to different combinations of oxide heterostructures and superlattices, such as $LaNiO_3$/LAO superlattices[277] and $GdTiO_3$/STO multilayers[278]. Band structures of other buried systems — such as dilute magnetic semiconductor (gallium, manganese) arsenic capped with amorphous arsenic layer[279] and spin injector $SiO_x$/EuO/silicon heterostructure[280] — were also measured by SX-ARPES and discussed in relation to the mechanism of ferromagnetism and performance of spintronic devices. Extending ARPES to the tender X-ray regime (HX-ARPES) enables probing of deeper interfaces. While some results have been demonstrated[281,282], it currently comes at a significant sacrifice of energy and momentum resolution. This makes SX-ARPES a more frequently used technique to directly visualize the band structure of buried systems, and further technological advancement of HX-ARPES to improve the energy and momentum is needed.

**Reproducibility and data deposition**

ARPES is an experimentally challenging technique, with extended varieties combining different light sources and detectors. Therefore, it is important to establish reproducibility, field standards, and data sharing for ARPES measurements.

**Issues with reproducibility**

Reproducibility of ARPES results requires careful consideration of some important issues such as the sample quality, photoexcitation, sample orientation, and other measurement conditions. Firstly, the sample quality and the surface cleanness are critical for obtaining high-quality ARPES data. Particular attention should be given to avoiding surface contaminations such as adsorption. Preparation procedures, aging of the open surface, temperature cycles and possible beam damage must be closely monitored and documented to guarantee data reproducibility. Secondly, careful determination of the sample's normal angles is required to ensure accurate conversion of the momentum. This is particularly important when comparing data measured under different experimental conditions, such as temperature, doping, light excitation, or comparing data measured on different samples. Thirdly, the photoexcitation characteristics are critical in ARPES measurements, which affects the probing depth and orbital sensitivity. Photon energy, polarization, experimental geometry, and sample orientation must all be accounted for when attempting to reproduce previous results. Due to the complexity of the ARPES measurements, results may need to be reproduced on multiple samples, or on different setups before publication. In addition, it is helpful to provide metadata annotation as extensively as possible for reproducibility among worldwide researchers.

**Repositories and data deposition**

To achieve sufficient interpretability, ARPES data must be expressed in terms of $E_B$ and $\mathbf{k}$. The $E_k$ measured by the analyzer must be converted to $E_B$ by shifting the energy scale. The most accurate approach consists of calibrating the Fermi energy of a reference metallic surface in electrical contact with the sample to zero. The emission angles are converted in momentum $\mathbf{k}$ by applying transformation matrices[117,283]. Current ARPES measured data are 2D or 3D intensity plots represented by color maps, therefore, EDCs or MDCs are needed when claiming critical observables such as an energy gap or bandwidth.

**Minimum reporting standards**

To ensure reproducibility, ARPES data must be supported by some critical information such as sample preparation method, overall energy resolution, photon energy and polarization, experimental geometry, sample temperature, and the facility used to perform the experiment. For TrARPES, it is also necessary to specify the overall time resolution, the pump fluence, and wavelength. For SpinARPES, it is important to report the detector spin-selectivity in the form of the effective Shermann function, while for NanoARPES the overall spatial resolution is an essential parameter.

**Limitations and optimizations**

**k_z broadening**

ARPES spectra inherently contain information of electronic states averaged over a finite $k_z$ window for 3D materials. This effect is called $k_z$ broadening[284,285]. The strength of $k_z$ broadening, $\Delta k_z$, is related to the short probing depth $\lambda$ via an equation $\lambda \cdot \Delta k_z \sim 1$ that stems from the Heisenberg uncertainty principle. For example, when surface-sensitive VUV photons with $\lambda \sim 5$ Å are used, $\Delta k_z \sim 0.2$ Å$^{-1}$, corresponding to several tens of percent of the bulk BZ length for typical materials, the ARPES line shape strongly suffers from $k_z$ broadening. Its effect must be carefully considered for states with a finite $k_z$ dispersion commonly observable in 3D and quasi-2D materials. On the other hand, if the target electronic states are purely 2D, as in the case of atomic-layer materials — graphene, TMDCs — and surface/interface states — topological Dirac-cone states, Rashba surface states — the $k_z$ broadening does not play a role, enabling straightforward interpretation of the spectral line shape. When a well-defined final state is not available due to the bandgap, the $k_z$ broadening can cover the full BZ length, known as a bandgap case[286]. This results in simultaneous observation of band dispersions predominantly from the $k_z = 0$ and $\pi$ planes because high symmetry planes contribute largely to the total spectral weight due to the high density of states at the band singularity points. A normal approach to reduce $k_z$ broadening is to increase $\lambda$, since the $k_z$ broadening is mainly governed by the uncertainty principle. According to the universal curve, low-energy VUV photons from laser or high energy SX and HX photons from synchrotron radiation are useful in this regard. However, the use of x-ray photons degrades the energy resolution and $k_{//}$ resolution. It is therefore useful to complementarily use the VUV and SX photons to elucidate intrinsic band dispersion and spectral line shape.

**Final-state effect**

The final state $\boldsymbol{n_f}, \boldsymbol{k_f}$ of photoexcited electrons can be affected by many factors that lead to broadening of the measured energy and k, called the final-state effect[54,287]. The final-state effect is sensitive to the choice of hv and polarization, which is not straightforward to calculate theoretically. A practical approach is to experimentally find the optimal photon energy for each compound by sweeping hv over a wide range and keeping in mind that an observed kz value varies at different photon energies. To detect all bands consisting of different electron orbitals, it is also helpful to use multiple independent light polarizations. It is also important to measure along high symmetry directions, and therefore careful alignment of the sample orientation is required, for example, by taking a Laue picture, an in situ low energy electron diffraction

(LEED) measurement, or a quick Fermi surface map by ARPES. Regarding the light polarization in real ARPES measurements, we note that it is difficult to obtain purely polarized light because the light is not at normal incidence with respect to the sample and the analyzer. A possible way to overcome this is to extract intensities for each component by adding or subtracting the ARPES signals at the original setting and those measured after rotating the sample in the azimuth direction by a particular angle depending on the crystal symmetry. This is especially useful for quantitative orbital analysis from circular dichroism[67].

**Space charge**

Photoemission measurements may be influenced by the space charge effect[288,289]. When the photoelectrons reach a sufficiently high density in the free space, they form a charge cloud and interact with each other, leading to trajectory changes and smearing the $E_k$. High concentrations of electrons can appear either at emission, just above the surface of the solid, or in the foci of the analyzer lenses. While vacuum space-charge effects can be minimized by continuous or quasi-continuous light sources, they represent a critical effect for time-resolved measurements[288].

In TrARPES, electrons are emitted within a narrow time window, achieving much higher densities at lower average photon fluxes. This is particularly dramatic in TOF momentum microscope analyzers that focus the photoelectron trajectories tightly at multiple points. Normally, space charge is limited by reducing the source flux. A flux scan is made to determine the acceptable compromise between statistical efficiency and space charge distortions[109,110]. New high-repetition-rate pulsed sources with high brilliance (HHG and FELs) are now coming online, which can greatly minimize the space charging issue.

When implemented in a pump-probe configuration, the pump pulse might create a dense cloud of photoelectrons, by multiphoton photoemission or by plasmonic hot-spot emission at sharp defects. This can give rise to complex kinematics. The pump-generated cloud is slower than the probe-generated one and they might cross each other for positive delays[61]. Experimental solutions by using a retarding electrostatic front lens are currently being developed to suppress secondary electrons and extend the space-charge limit[290].

**Outlook**

Over the past few decades, there have been major developments in ARPES. Higher spin-, spatial- and time-resolution have led to advances in understanding the fundamental electronic structure of quantum materials, such as superconductors, topological materials, 2D materials and heterostructures. Beyond the recent progress, further improvements of ARPES spectroscopy are expected, improving the energy, momentum and spatial resolution of HR-ARPES and NanoARPES through the upgrade of diffraction-limited synchrotron light sources. Major advances in TrARPES through high-repetition-rate laser and free-electron laser sources as well as tunable, ultra-strong pump field are also expected. SpinARPES data acquisition efficiency will be improved by multi-dimensional data acquisition capability.

Synchrotron light sources worldwide have been undergoing upgrades to 4th generation diffraction-limited operation. These light sources provide photon beams with significantly improved transverse coherence and higher intensity with a smaller beam size, allowing HR-ARPES with improved resolutions. In addition, the increasing coherent flux provides the possibility to achieve an even smaller spot size, down to sub-50 nm, for NanoARPES. This uses Fresnel zone plates to enable more measurements and data acquisition through wider sets of parameter space[291]. It can also be used to operate at higher photon energies and to achieve smaller focus spots, compensating for the decreased photoemission cross-sections that currently hinder the NanoARPES signal at photon energies $\geq$ 200 eV.

For TrARPES, major improvements to light sources, in particular the tunability and strength of the probe source, in addition to the pump fields, are also expected. TrARPES has recently reached an important milestone with the flourishing of high-repetition-rate harmonic sources that have unlocked access to the full span of the BZ[18,20,72,292]. These sources are likely to further develop over the next few years, achieving high degrees of polarization control, energy tunability, efficient and simpler driving laser sources, and preservation of temporal coherence between pump and probe pulses. A tunable probe source that allows the electron dynamics of 3D quantum materials to be probed was recently demonstrated[60,293]. High-repetition-rate free-electron lasers (FELs) such as the European XFEL[61], LCLS-II[294], and SHINE[295] will ramp up operations, allowing unprecedented usable photon flux to overcome the current statistical limitations of the technique. The advancement of both kinds of sources will enable the multi-modal ARPES, such as combining spin and spatial resolution to TrARPES.

Another development of TrARPES is the manipulation of the pump field. With the use of optical parametric amplifiers, the pump wavelength is tunable, allowing excitation of specific transitions. A change in perspective is obtained by using mid-IR or THz pulses. Tuning into optically-active phonon modes[296], the energy of the pump photons can be directly injected into the phononic system, leading to drastic phase transitions, such as light-induced superconductivity[297]. By accelerating the Dirac fermions using a THz light, the light-driven current can be tuned in a topological insulator surface[298]. Another interesting direction is Floquet engineering of quantum materials[299,300]. This can potentially enable dynamical engineering of the electronic structures and light-induced topological properties through the time-periodic light field.

Major improvements in SpinARPES can also be expected, particularly in data acquisition efficiency. SpinARPES instruments are typically only able to acquire data at single points in the momentum space at a time. This greatly reduces their overall efficiency compared to standard ARPES data acquisition. However, the impact of the technique continues to drive high demand for SpinARPES capability and continuous advancement of instrumentation. Advances have brought full imaging polarimeters to ARPES spectrometers[301,302], providing significant improvements to SpinARPES acquisition efficiency.

The emergence of multifunctional ARPES with higher resolutions provides a powerful tool to address leading scientific problems. For example, by combining ultrafast time resolution, high photon energy with large detectable $k$-space range, micrometer-scale beam spot and gating function, important information can be obtained to study correlated physics, such as the band structure engineering and correlation physics in MABLG systems via moiré-Floquet engineering by combining twistronics and light-induced Floquet engineering[303].

**Glossary**

**Bardeen-Cooper-Schrieffer superconductors:** Superconductors that can be well explained by Bardeen-Cooper-Schrieffer (BCS) theory developed in 1957, in which electron-phonon interaction plays a critical role in the superconductivity.

**Berry curvature:** A geometrical property of the electron wavefunctions in the parameter space, which is critical for the electronic properties of materials such as topological materials.

**Bloch state:** An electronic state of electrons in a periodic potential crystal, which is described as periodic wave function multiplied by a plane wave with specific wavevector.

**Brillouin zone:** A primitive cell in reciprocal ($k$) space. The first Brillouin zone consists of a set of points in $k$-space. This set of points that can be reached from the origin without crossing any Bragg plane.

**Cooper pairing:** Pairing of two electrons at low temperature to form a bound state with the character of a boson, the condensation of which is the foundation for the BCS superconductivity.

**Energy distribution curve:** A common lower dimensionality (1D) subset of ARPES data used in data analysis, consisting of the photoemission intensity distribution versus energy at a fixed momentum value.

**Excitons:** Bound states of electrons and holes in an insulator or semiconductor. These many-body states, bound by electrostatic Coulomb attraction, can be treated as single quasiparticles.

**Fermi-Dirac distribution function:** The Fermi-Dirac distribution function, also called the Fermi function. For a certain temperature, it provides the energy level occupancy probability by fermions.

**Free-electron approximation:** Assumption in the context of photoemission that the photoelectrons obey the parabolic dispersion relationship between energy and momentum of free electrons. This ignores long-range interactions between the emitted electron and the solid.

**Hemispherical analyzer:** An electron kinetic energy analyzer that produces circular orbits with hemispherical capacitors. The radii of the electron orbits depend on their kinetic energy, mapping it linearly to a position at the exit of the hemispheres.

**Inelastic collisions:** A scattering process in which the energy of the individual particles is not conserved.

**Inelastic mean free path:** Mean distance that electrons can travel before suffering an inelastic collision process. It is material and energy dependent.

**Kramers-Kronig relation:** A mathematical relation between the real and imaginary parts of a complex function like the energy-dependent electron self-energy.

**Magnons:** Quasiparticles describing a collective excitation of spin order, i.e. a quantization of a spin-wave.

**Momentum distribution curve:** A common lower dimensionality (1D) subset of ARPES data used in data analysis, consisting of the photoemission intensity distribution versus momentum at a fixed energy value.

**Photoelectric effect:** Release of electrons from a material surface upon irradiation with light.

**Phonons:** Quasiparticles describing collective excitations of atomic vibrations in a crystal.

**Photoelectrons:** Electrons released from the material surface by the photoelectric effect.

**Plasmons:** Quantization of collective oscillations of charge carriers in a solid, arising when electromagnetic fields act on a conducting surface.

**Plasmarons:** Composite quasiparticles formed by charges and plasmons, which can modify the electronic dispersion.

**Pump-probe measurements:** Stroboscopic method to observe non-equilibrium electronic distributions and their dynamics. A pulsed pump beam excites the electrons, followed by a delayed pulse that produces photoemission. This routine is repeated for multiple pump-probe delays, mapping the dynamical evolution.

**Quasiparticle:** A quasiparticle is an emergent quantum excitation of a many-body system that can be treated as a non-interacting particle, given a suitable renormalization of its properties.

**Time-of-flight (TOF) analyzer:** An electron kinetic energy analyzer. Kinetic energy is resolved by measuring the electron drift time through a long field-free section of the electron optics. Shortly pulsed sources are required to time the electron start, along with fast electron detectors to time the electron arrival.

**Topological insulators:** Topological insulators are materials with a bulk band gap, like ordinary insulators, but on their edge or surface they also have protected conducting states, known as topological states. Topological states are induced by the combination of time-reversal symmetry and spin-orbit coupling.

**Wannier function:** Element of a complete set of localized orthogonal functions. They are a convenient representation when describing highly localized wavefunctions or local properties of a wavefunction in a solid.

**Work function:** Minimum energy required to remove an electron from a solid surface to a distance such that all interactions with the solid are negligible.

**Acknowledgment:**

We thank Hayley Wallen for the help during the manuscript preparation. Hongyun Zhang & Shuyun Zhou are supported by the National Key R & D Program of China (Grant No. 2021YFA1400100, 2020YFA0308800), National Natural Science Foundation of China (Grant No. 11725418), Tohoku-Tsinghua Collaborative Research Fund. Tommaso Pincelli is supported by the Alexander von Humboldt Stiftung with the Alexander von Humboldt Fellowship. Ralph Ernstorfer is supported by the Max Planck Society, the European Research Council (ERC) under the European Union's Horizon 2020 research and innovation program (Grant No. ERC-2015-CoG-682843, H2020-FETOPEN-2018-2019-2020-01 (OPTOLogic—grant agreement No. 899794)). Chris Jozwiak is supported by the Advanced Light Source, a U.S. DOE Office of Science User Facility under Contract No. DE-AC02-05CH11231. Takeshi Kondo is supported by the JSPS KAKENHI (Grant No. JP21H04439, JP19H00651), by MEXT Q-LEAP (Grant No. JPMXS0118068681), and by MEXT as "Program for Promoting Researches on the Supercomputer Fugaku" (Basic Science for Emergence and Functionality in Quantum Matter Innovative Strongly Correlated Electron Science by Integration of "Fugaku" and Frontier Experiments) (Project ID: hp200132). Takafumi Sato is supported by JST-CREST (No. JPMJCR18T1) and JSPS (KAKENHI No. 21H04435).

**Competing interests: The authors declare no competing interests**